\documentclass[twocolumn,showpacs,preprintnumbers,amsmath,amssymb]{revtex4-1}

\usepackage{graphicx}
\usepackage{dcolumn}
\usepackage{bm}
\usepackage{array}
\usepackage{tabularx}
\newcolumntype{L}[1]{>{\raggedright\arraybackslash}p{#1}}
\newcolumntype{C}[1]{>{\centering\arraybackslash}p{#1}}
\newcolumntype{R}[1]{>{\raggedleft\arraybackslash}p{#1}}

\begin{document}

\preprint{gxz sign reversal paper PRB vfinal}

\title{Electrical control of the sign of the $g$-factor in a GaAs hole quantum point contact}

\author{A. Srinivasan$^{1}$, K. L. Hudson$^{1}$, D. S. Miserev$^{1}$, L. A. Yeoh$^{1}$, O. Klochan$^{1}$, K. Muraki$^{2}$, Y. Hirayama$^{3}$, O. P. Sushkov$^{1}$}
\author{A. R. Hamilton$^{1}$}
\email{Alex.Hamilton@unsw.edu.au}
\affiliation{$^{1}$School of Physics, University of New South Wales, Sydney NSW 2052, Australia}
\affiliation{$^{2}$NTT Basic Research Laboratories, NTT corporation, Atsugi-shi, Kanagawa 243-0198, Japan}
\affiliation{$^{3}$Graduate School of Science, Tohoku University, Sendai-shi, Miyagi 980-8578 Japan}

\date{\today}

\begin{abstract}
Zeeman splitting of 1D hole subbands is investigated in quantum point contacts (QPCs) fabricated on a (311) oriented GaAs-AlGaAs heterostructure.  Transport measurements can determine the magnitude of the $g$-factor, but cannot usually determine the sign.  Here we use a combination of tilted fields and a unique off-diagonal element in the hole $g$-tensor to directly detect the sign of $g^{*}$.  We are able to tune not only the magnitude, but also the sign of the $g$-factor by electrical means, which is of interest for spintronics applications.  Furthermore, we show theoretically that the resulting behaviour of $g^{*}$ can be explained by the momentum dependence of the spin-orbit interaction.
\end{abstract}

\maketitle

Electrical manipulation of spin is the underlying principal of many proposed spintronic and quantum computing device architectures \cite{DattaApl90, LossPRA98, WolfSci01, Awsbook02}.  In particular, electrical control of the effective Land\'{e} $g$-factor in semiconductor nanostructures has been a major focus of recent research, with theoretical investigations predicting strong $g^{*}$ tunability in both magnitude and sign \cite{PradoPRB04, KuglerPRB09, AndlauerPRB09}.  The ability to invert the sign of the $g$-factor and tune the system through a state of zero spin polarisation ($g^{*} = 0$) could be a valuable asset in engineering solid-state spin devices \cite{SalisNat01, KatoSci03, BennettNcom13}.  

In this regard, quantum confined hole systems in GaAs are prime candidates due to the strong coupling between spin and orbital motion in the valence band \cite{WinklerBook03}.  The spin 3/2 nature of valence band holes in GaAs leads to several unique properties such as a tensor structure of $g^{*}$ with large anisotropy between all three spatial directions \cite{KesterenPRB90, WinklerPRL00}, and tunability of the $g$-factor across orders of magnitude \cite{DanneauPRL06, SrinivasanNL13, NichelePRL14}.

Previous studies of the $g$-factor of quantum confined holes revealed a non-monotonic dependance of $|g^{*}|$ on the gate bias, suggestive of a change in sign of $g^{*}$ \cite{KuglerPRB09, KlochanNJP09}.  However, these studies could not directly detect the sign of $g^{*}$, only its magnitude.  In this work, we utilise a novel approach to directly detect the sign of $g^{*}$ by exploiting a unique property of the (311) GaAs hole $g$-tensor, and demonstrate a gate-controlled sign change of $g^{*}$ in a hole quantum point contact (QPC) on (311) GaAs.  

We also introduce a theoretical model showing that the observed sign reversal of $g^{*}$ arises from the in-plane momentum dependence of the spin-orbit interaction in the valence band.  Typically it is not possible to experimentally probe the directional $k$-dependence of the 2D hole $g$-tensor, since transport measurements represent an average over all $k$-states at the Fermi surface.  However, by using an electrostatically controlled QPC fabricated along particular in-plane directions of a 2D hole system, we can perform a direct spectroscopic measurement of $g^{*}$, and investigate its dependence on the magnitude and direction of the in-plane momentum \cite{DanneauPRL06, KlochanNJP09, ChenNJP10}.

The device used in this work was fabricated from a (311)A-oriented heterostructure, in which a 2D hole system is induced at an AlGaAs/GaAs interface by applying a negative voltage (-0.7V) to a heavily p-doped cap layer \cite{ClarkeJAP06}. The peak 2D hole mobility was $\mu = 6.0 \times 10^5$~cm$^{2}$ V$^{-1}$s$^{-1}$ at a density $p=1.3 \times 10^{11}$~cm$^{-2}$ and temperature T = 40 mK. The 2D holes are further confined using a split-gate geometry, to two short one-dimensional (1D) channels or quantum point contacts (QPCs) - see Fig.1a.  The two orthogonal 400nm long 1D channels, oriented along the [$\overline{2}33$] and [$01\overline{1}$] crystal directions (which we label QPC$[\overline{2}33]$ and QPC$[01\overline{1}]$ respectively), were defined by electron-beam lithography and shallow wet etching of the cap layer. Measurements were carried out in a dilution refrigerator, with a base temperature below 40mK, using standard ac lock-in techniques with a $100\mu V$ excitation at 31Hz.  A three-axis vector magnet was used to independently control all three components of the magnetic field, eliminating the need to thermally cycle the device.  The fields were applied along $[\overline{2}33]$ and $[311]$ as shown by the schematic in Fig.1b.

\begin{figure}
\includegraphics[width=0.99\linewidth]{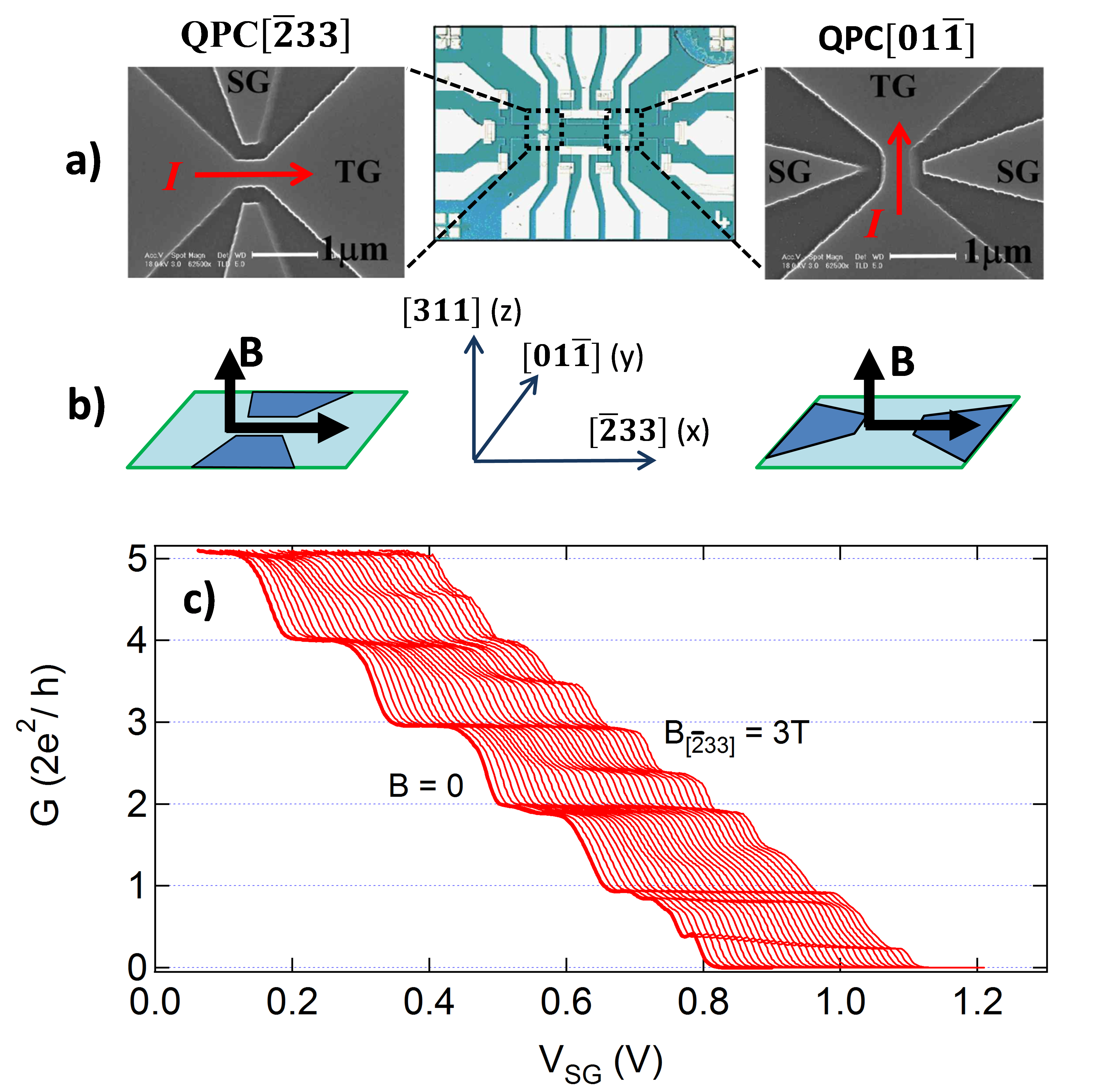}
\caption{ (a) The centre panel shows an optical image of the device fabricated on (311)A GaAs. The left and right panels show electron micrographs of the two orthogonal QPCs along the $[\overline{2}33]$ and $[01\overline{1}]$ directions respectively (red arrows indicate current direction). (b) Schematic diagram showing the orientation of the magnetic fields used in this experiment with respect to the two QPCs.
(c) The conductance G versus $V_{SG}$ for QPC$[\overline{2}33]$ in a magnetic field applied along $[\overline{2}33]$, showing characteristic 1D conductance plateaus at B = 0, which evolve into spin resolved half-plateaus at finite fields (traces offset for clarity).}
\end{figure}

Fig.1c shows the conductance as QPC$[\overline{2}33]$ is pinched off, revealing clean 1D conductance plateaus in units of $2e^2/h$ at B = 0, which evolve to spin resolved half plateaus when a magnetic field was applied along the in-plane $[\overline{2}33]$ direction.  The $g$-factor was extracted by measuring the Zeeman splitting in gate voltage $\Delta V_{SG}(B)$, which is then converted to a Zeeman energy splitting $\Delta E_{Z}(B)$ using the well known source drain bias spectroscopy technique \cite{PatelPRB91} (see Supplemental Material \cite{supp1} section 1).

Figs. 2a and 2b show the Zeeman splitting of the 1D subbands in the two orthogonal QPCs with a magnetic field $B_{[\overline{2}33]}$ applied.  The greyscale plots show the transconductance $\partial G/\partial V_{SG}$, with the dark regions corresponding to the risers between plateaus in Fig. 1c, hence marking the 1D subband edges.  

\begin{figure}
\includegraphics[width=0.99\linewidth]{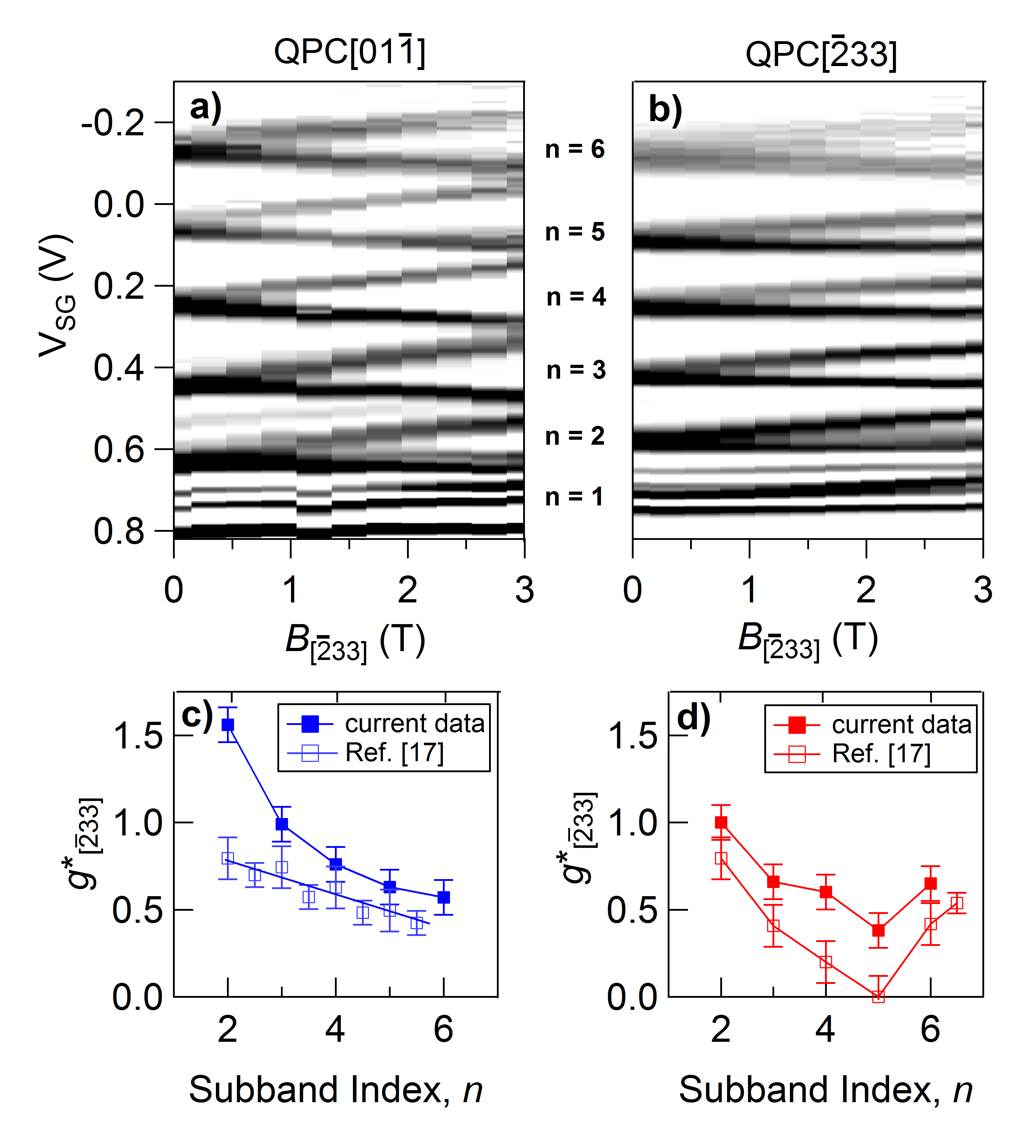}
\caption{Top panels show Zeeman splitting in an in-plane magnetic field applied along the $[\overline{2}33]$ direction for a) QPC$[01\overline{1}]$ and b) QPC$[\overline{2}33]$. Greyscale plots of the transconductance $\partial G/\partial V_{SG}$ are shown, where the dark regions represent the 1D subband edges.  Bottom panels show the effective $g$-factors measured for (c) QPC$[01\overline{1}]$ and (d) QPC$[\overline{2}33]$. QPC$[\overline{2}33]$ shows a non-monotonic trend, indicative of a sign change of $g^{*}_{[\overline{2}33]}$ at $n = 5$.}
\end{figure}

For both QPCs there is a clear linear Zeeman splitting of the 1D states, from which we extract the $g$-factor. The measured $g^{*}_{[\overline{2}33]}$ for QPC$[01\overline{1}]$ is plotted in Fig. 2c along with earlier data from Ref. \cite{KlochanNJP09} taken at a higher 2D hole density.  In both cases, $g^{*}_{[\overline{2}33]}$ shows a monotonic decrease with increasing subband index $n$.  The equivalent $g$-factor for QPC$[\overline{2}33]$ is shown in Fig. 2d, and we again show earlier data taken at a higher density \cite{KlochanNJP09}.  In contrast to QPC$[01\overline{1}]$, QPC$[\overline{2}33]$ shows a non-monotonic evolution of $g^{*}_{[\overline{2}33]}$ as a function of subband index, with a clear minimum at $n = 5$.  This marked difference in the $g$-factor for orthogonal current directions is due to a combination of the crystallographic anisotropy in the (311) surface and the in-plane momentum dependence of $g^{*}$, as shown later.

We now use a novel approach to prove that the trend observed in Fig. 2d is due to a sign change of the in-plane $g$-factor $g^{*}_{[\overline{2}33]}$, as the 1D channel is tuned from the 2D to the 1D limit.  Although the observed non-monotonic trend of $g^{*}_{[\overline{2}33]}$ is suggestive of a sign reversal, these measurements alone cannot determine the sign of $g^{*}$.  In the following section, we show that the sign of $g^{*}$ can be explicitly extracted by simultaneously applying orthogonal magnetic fields to exploit an unusual property of the (311) hole $g$-tensor:  Uniquely to (311) oriented GaAs 2D systems, theory \cite{WinklerSST08} and experiment \cite{YeohPRL14} have shown that when a field is applied along the in-plane $[\overline{2}33]$ direction, in addition to an in-plane polarisation with $g$-factor $g_{xx}$, there exists an anomalous out-of-plane polarisation due to an off-diagonal term $g_{xz}$ in the $g$-tensor.  The Hamiltonian describing the Zeeman term for 2D heavy holes in (311) GaAs is then:
\begin{multline}
H~=~\frac{\mu_{B}}{2}((g_{xx}B_{x}\sigma_{x}) + (g_{xz}B_{x}\sigma_{z}) + (g_{zx}B_{z}\sigma_{x})\\
+ (g_{yy}B_{y}\sigma_{y}) + (g_{zz}B_{z}\sigma_{z}))
\end{multline}
where $x$,$y$ and $z$ refer to the $[\overline{2}33]$, $[01\overline{1}]$ and [311] directions respectively, with theoretical 2D values $g_{xx} = g_{yy} = -0.16$, $g_{xz} = 0.65$, $g_{zz} = 7.2$ \cite{WinklerSST08} and $g_{zx} \simeq 0$ \cite{YeohPRL14}.  With the magnetic field applied along $[01\overline{1}]$, the Zeeman splitting is $\Delta E_{Z} = g^{*}_{[01\overline{1}]}\mu_{B}B_{[01\overline{1}]}$, where $g^{*}_{[01\overline{1}]}$ is simply the isotropic component of the $g$-tensor $g_{yy}$.  However, when the field is applied along $[\overline{2}33]$, the Zeeman splitting is $\Delta E_{Z} = g^{*}_{[\overline{2}33]}\mu_{B}B_{[\overline{2}33]}$, where $|g^{*}_{[\overline{2}33]}| = \sqrt{{g_{xx}}^2 + {g_{xz}}^2}$.  

If combined magnetic fields are applied both along the in-plane $[\overline{2}33]$ and out-of-plane [311] directions, the total Zeeman splitting measured in experiment is:
\begin{multline}
\Delta E_{Z}^2~=~{(g_{xx}\mu_{B}B_{[\overline{2}33]})}^2 + {(g_{xz}\mu_{B}B_{[\overline{2}33]} + g_{zz}\mu_{B}B_{[311]})}^2\\
\end{multline}
The resulting Zeeman spliting is unusual in that it is sensitive to the relative signs of the $g_{xz}$ and $g_{zz}$ terms: If both $g_{xz}B_{[\overline{2}33]}$ and $g_{zz}B_{[311]}$ have the same sign, the total Zeeman splitting is large. However, if one of the two terms is negative, the total Zeeman splitting is suppressed.  Therefore, applying both $B_{[\overline{2}33]}$ and $B_{[311]}$ simultaneously allows the relative signs of $g_{xz}$ and $g_{zz}$ to be extracted.  

To check if there is a sign change of $g^{*}_{[\overline{2}33]}$ as suggested by Fig. 2d, we again measure the Zeeman splitting of 1D subbands as a function of $B_{[\overline{2}33]}$ but now apply an additional fixed magnetic field along the out-of-plane [311] direction.  The magnitude of the total Zeeman splitting depends on the relative signs of the $g_{xz}B_{[\overline{2}33]}$ and $g_{zz}B_{[311]}$ terms in eqn. 2, resulting in an asymmetry in the Zeeman splitting around $B_{[\overline{2}33]} = 0$.  Crucially, if the sign of $g_{xz}$ changes with respect to $g_{zz}$, the asymmetry in the Zeeman splitting as a function of $B_{[\overline{2}33]}$ should reverse, providing direct proof of a sign reversal \cite{signofgxx}. 


Turning to the experimental results, Fig. 3 shows the Zeeman splitting of both QPC$[01\overline{1}]$ and QPC$[\overline{2}33]$ in combined magnetic fields applied in and out of the plane.  When a fixed out-of-plane field $B_{[311]} = 0.2T$ is introduced (Figs. 3a and 3b), the data becomes asymmetric around $B_{[\overline{2}33]} = 0$.  We note that for 1D holes on the high symmetry (100) plane, the data is always symmetric even in combined magnetic fields, due to the absence of the off-diagonal $g_{xz}$ term (see supplemental material \cite{supp1} section 2).

\begin{figure}
\includegraphics[width=0.99\linewidth]{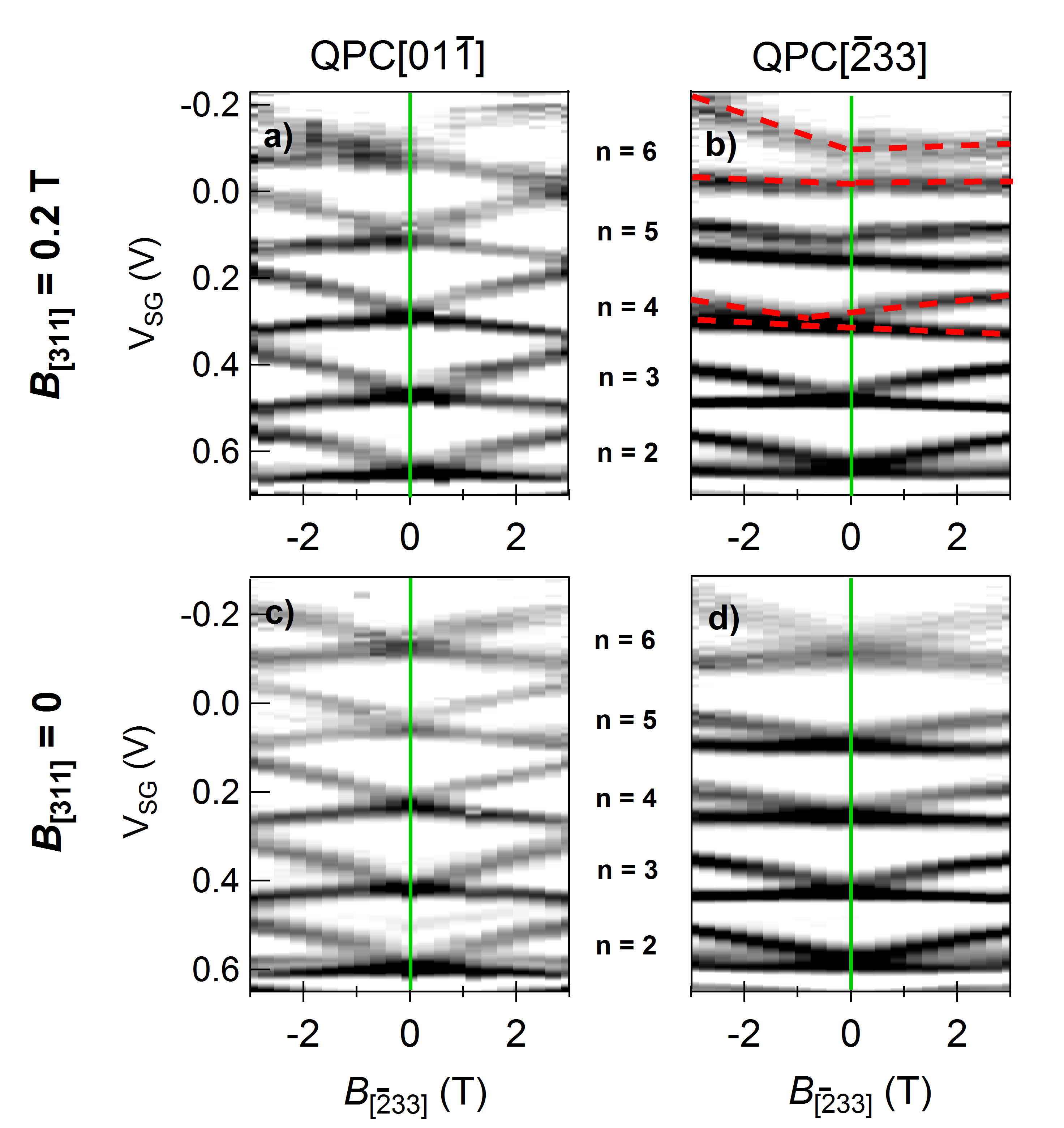}
\caption{Greyscale transconductance plot of the Zeeman splitting for both QPCs as a function of $B_{[\overline{2}33]}$, for $B_{[311]} = 0.2 T$ (a and b) and $B_{[311]} = 0$ (c and d). The dark regions represent the 1D subband edges.  The out-of-plane field $B_{[311]}$ causes an asymmetry around $B_{[\overline{2}33]} = 0$ due to the interplay between $g_{zz}$ and $g_{xz}$.  For subband 6, QPC$[\overline{2}33]$ shows a reversal of asymmetry due to the sign change of $g_{xz}$, as indicated by the red dashed lines.}
\end{figure}

Starting with QPC$[01\overline{1}]$ (Fig. 3a), the lower subbands do not appear to show any asymmetry in the combined fields, suggesting that the cancellation/addition of $g_{zz}$ and $g_{xz}$ is minimal (this is due to the fact that $g_{zz}$ is small for low subbands - see supplemental material \cite{supp1} section 3).  However, for subbands 5 and 6, the asymmetry around $B_{[\overline{2}33]} = 0$ becomes increasingly apparent as $g_{zz}$ becomes large.  Subband 6 clearly shows a strong Zeeman splitting for $B_{[\overline{2}33]} > 0$, and a relatively weak splitting for $B_{[\overline{2}33]} < 0$.  This confirms the predicted effect due to the competition between the $g_{zz}$ and $g_{xz}$ terms in eqn. 2.  
In the case of QPC$[\overline{2}33]$ (Fig. 3b), the asymmetry of the Zeeman splitting around $B_{[\overline{2}33]} = 0$ again increases with subband index.  However, the most significant aspect of the data is that the asymmetry is \textit{reversed} for subband 6, which can only occur if $g_{xz}$ has changed sign between $n = 5$ and $n = 6$ \cite{gzzsign}.  This is consistent with the data in Fig. 2d, where there is a clear minimum around $n = 5$.

In order to confirm that the asymmetry in the Zeeman splitting is caused by the combination of magnetic fields, we also show the Zeeman splitting as a function of $B_{[\overline{2}33]}$, with $B_{[311]}$ = 0 (Figs. 3c and 3d).  In this case, the $g_{zz}B_{[311]}$ term in eqn. 2 becomes zero, so the Zeeman splitting is simply $\Delta E_{Z}^2~=~(g_{xx}^2 + g_{xz}^2)~B_{[\overline{2}33]}^2 = g^{*2}_{[\overline{2}33]}~B_{[\overline{2}33]}^2$, resulting in a symmetric evolution of the subbands either side of $B_{[\overline{2}33]} = 0$.  The symmetry is clearly evident for both QPCs in Figs. 3c and 3d. 

We now turn to the question of what is causing the sign change of $g_{xz}$ for QPC$[\overline{2}33]$, and show theoretically that the data can be well explained by the dependence of the 2D $g$-factor on the in-plane momentum.  The 1D subband index effectively corresponds to quantised values of the in-plane momentum $\langle p_{\parallel}^2 \rangle$: In the 1D region, $\langle p_{\parallel}^2 \rangle$ is determined by the difference between the Fermi energy $E_{F}$ in the 2D reservoirs and the top of the saddle point potential created by the QPC gates \cite{Buttiker, ChenNJP10}.  In the 1D limit at $n$ = 1, the saddle point is high in energy and $\langle p_{\parallel}^2 \rangle$ is small. As the subband index increases, the saddle point decreases in energy so $\langle p_{\parallel}^2 \rangle$ also grows larger and eventually saturates at $\langle p_{\parallel}^2 \rangle = p_{F}^2$.  Hence, by tuning the 1D subband index, we are effectively probing the effects of finite momentum on $g^{*}$.

We now analyse how $g_{xz}$ should depend on the in-plane momentum and directly relate this to the measurements of $g_{xz}$ vs $n$ for both QPCs.  We begin with the Luttinger Hamiltonian and take into account both the axial and cubic terms corresponding to the crystallographic anisotropy of the (311) surface.  The 2D ($z$) confinement at the GaAs-AlGaAs interface, is taken as a triangular potential, and is assumed to be far greater than the in-plane ($x,y$) confinement due to the QPC, meaning we treat the hole system as quasi-2D in the ($x,y$)-plane with strong quantisation in the z-direction.  The in-plane momentum is then taken into account using perturbation theory with the parameter $\langle p_{\parallel}^2 \rangle / \langle p_{z}^2 \rangle$, where $\langle p_{\parallel}^2 \rangle = (\langle p_{x}^2 \rangle, \langle p_{y}^2 \rangle)$.  We consider a magnetic field applied in the $[\overline{2}33] (x)$ direction, and derive an expression for $g_{xz}$ as a function of $\langle p_{x}^2 \rangle$ and $\langle p_{y}^2 \rangle$ (see supplemental material section 5 for full derivation \cite{supp1}): 
\begin{multline}
g_{xz}~=~0.39 - C_{1}\frac{\langle p_{x}^2 \rangle}{\langle p_{z}^2 \rangle} - C_{2}\frac{\langle p_{y}^2 \rangle}{\langle p_{z}^2 \rangle} - C_{3}\frac{\langle p_{x}^2 \rangle - \langle p_{y}^2 \rangle}{\langle p_{z}^2 \rangle}
\end{multline}
The constants $C_{1}, C_{2}$ and $C_{3}$ depend on band structure parameters and the 2D confinement potential.  We have also included the Dresselhaus interaction which suppresses the $g$-factor by $\simeq 40\%$.  We note that the Rashba interaction makes a negligible contribution to $g^{*}$ \cite{supp1}.

\begin{figure}
\includegraphics[width=0.99\linewidth]{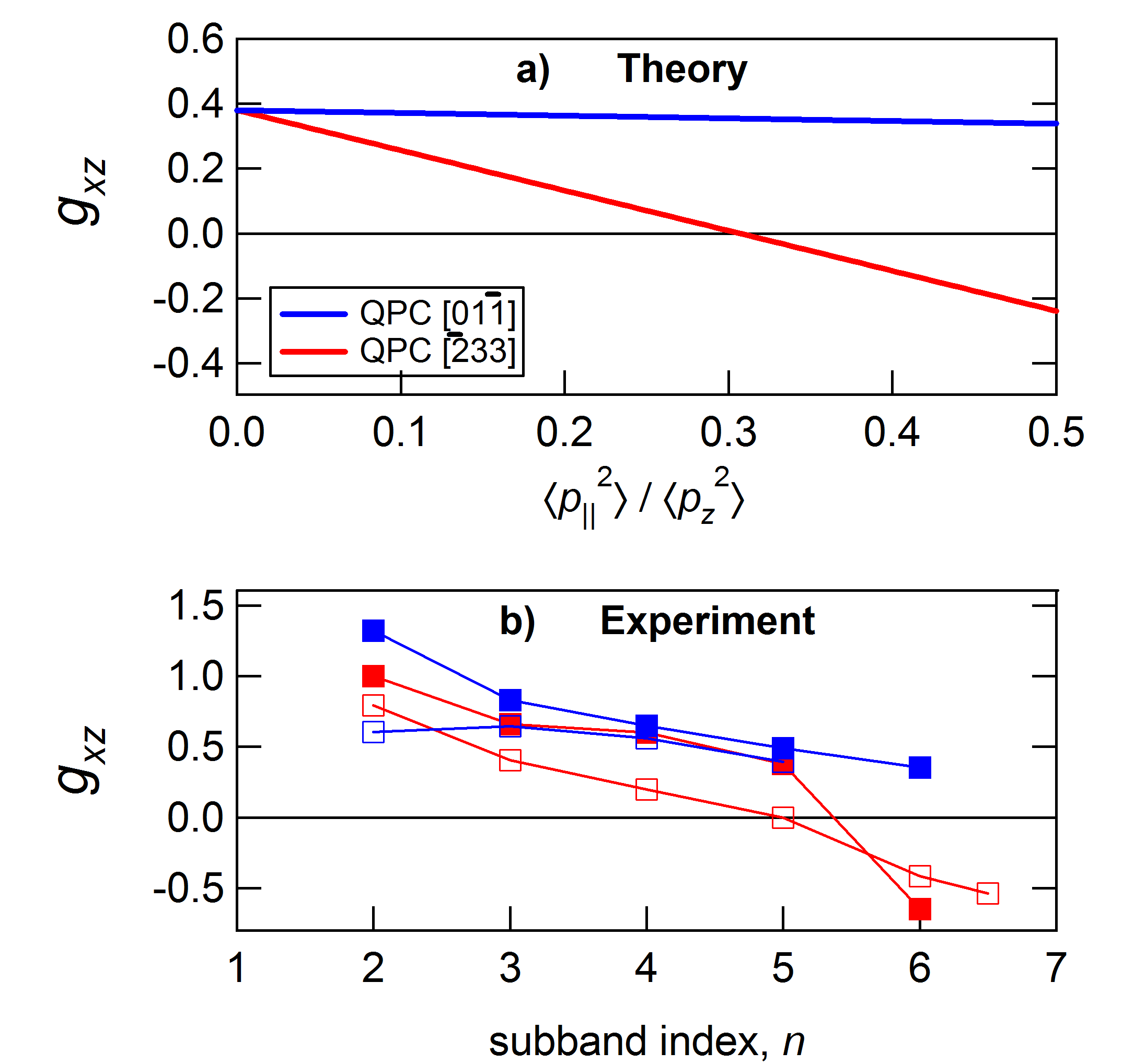}
\caption{(a) Theoretically predicted dependence of $g_{xz}$ on the in-plane momentum for QPC$[01\overline{1}]$ (blue) and QPC$[\overline{2}33]$ (red). $g_{xz}$ for QPC$[\overline{2}33]$ is strongly suppressed as a function of $\langle p_{\parallel}^2 \rangle$ and changes sign.  (b) Experimentally measured $g_{xz}$ versus subband index for both QPCs.  Solid squares correspond to data from this experiment, and open squares are data from ref. \cite{KlochanNJP09} at a higher 2D density.}
\end{figure}

The QPC confinement is taken into account as follows: For QPC$[\overline{2}33]$, the current is along the $x$ direction, so $\langle p_{x}^2 \rangle = 0$ since the spin splitting is measured at the subband edge, and $\langle p_{y}^2 \rangle$ takes quantised values corresponding to the 1D subbands.  Conversely, for the orthogonal QPC$[01\overline{1}]$, $\langle p_{y}^2 \rangle = 0$ and $\langle p_{x}^2 \rangle$ takes quantised values.  In Fig.4a, the theoretically calculated $g_{xz}$ is plotted as a function of $\langle p_{\parallel}^2 \rangle / \langle p_{z}^2 \rangle$.  The blue trace shows QPC$[01\overline{1}]$ with $\langle p_{\parallel}^2 \rangle = \langle p_{x}^2 \rangle$, and the red trace shows QPC$[\overline{2}33]$ with $\langle p_{\parallel}^2 \rangle = \langle p_{y}^2 \rangle$. Due to the differing dependence of $g_{xz}$ on $\langle p_{x}^2 \rangle$ and $\langle p_{y}^2 \rangle$ in eqn. 3 (originating from the crystallographic anisotropy of the (311) surface), the two orthogonal QPCs show strikingly different behaviour.  $g_{xz}$ for QPC$[01\overline{1}]$ is positive and decreases slightly with increasing $\langle p_{\parallel}^2 \rangle / \langle p_{z}^2 \rangle$ (and subband index), whereas $g_{xz}$ for QPC$[\overline{2}33]$ starts at a positive value but changes sign at larger  $\langle p_{\parallel}^2 \rangle / \langle p_{z}^2 \rangle$.  

The experimentally measured $g_{xz}$ for both QPCs, obtained from $g^{*}_{[\overline{2}33]}$ in Figs.2c and 2d, ($g_{xz} = \sqrt{g^{2}_{[\overline{2}33]} - g_{xx}^{2}} = \sqrt{g^{2}_{[\overline{2}33]} - g^{2}_{[01\overline{1}]}}$ - see section 4 of supplemental material \cite{supp1}) is plotted in Fig.4b.  The data shows good agreement with the theory, with $g_{xz}$ for QPC$[01\overline{1}]$ decreasing slightly as the in-plane momentum increases.  Meanwhile, $g_{xz}$ for QPC$[\overline{2}33]$ decreases strongly and changes sign around $n=5$.  In the limit of the largest measurable subband - subband 7, we use the known 2D density and confinement potential to numerically estimate the quantity $\langle p_{\parallel}^2 \rangle / \langle p_{z}^2 \rangle$ giving $\langle p_{\parallel}^2 \rangle / \langle p_{z}^2 \rangle \simeq 0.2$.  The sign change (at n=5) should therefore occur at $\langle p_{\parallel}^2 \rangle / \langle p_{z}^2 \rangle \lesssim 0.2$, which is reasonably close to the theoretically predicted value of $\langle p_{\parallel}^2 \rangle / \langle p_{z}^2 \rangle = 0.3$.  This small discrepancy may be due to the fact that the theory does not take into account the effects of 1D quantisation, which may alter the confinement parameters used to derive eqn. 3.  Nevertheless, the behaviour we observe for $g_{xz}$ in both QPCs is qualitatively consistent with that predicted by theory.

Finally we note that although the form of $g_{xz}$ obtained from the theory agrees well with experiment, a quantitative comparison shows that the range of $g_{xz}$ measured experimentally ($-0.65 < g_{xz} < 1.5$) is larger than that predicted by theory ($-0.3 < g_{xz} < 0.4$).  This enhancement of the $g$-factor in experiment may be attributed to many-body interactions (not included in the theoretical calculation), previously observed in both 1D electron and hole systems \cite{ThomasPRL96, DaneshvarPRB97}.  

In conclusion, Zeeman splitting measurements of 1D subbands were carried out for two orthogonal hole QPCs on (311)A GaAs.  Due to the low symmetry of the (311) surface, the total Zeeman splitting in combined fields becomes sensitive to the sign of different components of the $g$-tensor.  In this way, we are able to prove that $g_{xz}$ changes sign when the 1D channel is oriented along $[\overline{2}33]$, consistent with a theoretical model of $g^{*}$ versus in-plane momentum.  Our experimental results shed light on the complex spin physics of holes, and demonstrates gate-controlled tuning, not only of the magnitude but also the sign, of the $g$-factor, which is desirable for spintronics applications.  

\begin{acknowledgments}
The authors acknowledge the late J. Cochrane for technical support, and thank T. Li and U. Z\"{u}licke for enlightening discussions.  YH acknowledges support by KAKENHI Grant No. 26287059. This work was supported by the Australian Research Council under the DP scheme, and was performed in part using facilities of the NSW Node of the Australian National Fabrication Facility.
\end{acknowledgments}


\begin{thebibliography}
\bibitem{signofgxx} We note that while the Zeeman splitting in combined fields is sensitive to a relative sign change between $g_{xz}$ and $g_{zz}$, it cannot discern a sign change in the isotropic component of the $g$-tensor $g_{xx}$, as is evident from eqn 2.
\end{thebibliography}
\end{document}